\documentclass[twocolumn,journal]{IEEEtran}
\usepackage[T1]{fontenc}
\usepackage[latin9]{inputenc}
\usepackage{float}
\usepackage{amsmath}
\usepackage{amssymb}
\usepackage{stackrel}
\usepackage{graphicx}
\usepackage[unicode=true,
 bookmarks=true,bookmarksnumbered=true,bookmarksopen=true,bookmarksopenlevel=1,
 breaklinks=false,pdfborder={0 0 0},pdfborderstyle={},backref=false,colorlinks=false]
 {hyperref}
\hypersetup{pdftitle={Your Title},
 pdfauthor={Your Name},
 pdfpagelayout=OneColumn, pdfnewwindow=true, pdfstartview=XYZ, plainpages=false}

\makeatletter

\floatstyle{ruled}
\newfloat{algorithm}{tbp}{loa}
\providecommand{\algorithmname}{Algorithm}
\floatname{algorithm}{\protect\algorithmname}

\usepackage[caption=false,font=footnotesize]{subfig}

\makeatother

\begin{document}
 
\title{Using Loaded N-port Structures\\
 to Achieve the Continuous-Space \\
Electromagnetic Channel Capacity Bound}
\author{Zixiang Han,~ Shanpu Shen,~ Yujie Zhang,~ Shiwen Tang,~ Chi-Yuk
Chiu, and~Ross Murch\thanks{This work was supported by the Hong Kong research councils Collaborative
Research Fund (CRF) grant C6012-20G. This paper has been submitted
to IEEE Transactions on Wireless Communications. \textit{(Corresponding
author: Shanpu Shen.)}}\thanks{Z. Han, S. Shen, Y. Zhang, S. Tang, and C. Y. Chiu are with the Department
of Electronic and Computer Engineering, The Hong Kong University of
Science and Technology, Clear Water Bay, Kowloon, Hong Kong (e-mail:
sshenaa@connect.ust.hk).}\thanks{R. Murch is with the Department of Electronic and Computer Engineering
and the Institute for Advanced Study, The Hong Kong University of
Science and Technology, Clear Water Bay, Kowloon, Hong Kong.}}
\maketitle
\begin{abstract}
A method for achieving the continuous-space electromagnetic channel
capacity bound using loaded $N$-port structures is described. It
is relevant for the design of compact multiple-input multiple-output
(MIMO) antennas that can achieve channel capacity bounds when constrained
by size. The method is not restricted to a specific antenna configuration
and a closed-form expression for the channel capacity limits are provided
with various constraints. Furthermore, using loaded $N$-port structures
to represent arbitrary antenna geometries, an efficient optimization
approach is proposed for finding the optimum MIMO antenna design that
achieves the channel capacity bounds. Simulation results of the channel
capacity bounds achieved using our MIMO antenna design with one square
wavelength size are provided. These show that at least 18 ports can
be supported in one square wavelength and achieve the continuous-space
electromagnetic channel capacity bound. The results demonstrate that
our method can link continuous-space electromagnetic channel capacity
bounds to MIMO antenna design.
\end{abstract}

\begin{IEEEkeywords}
Beamspace, channel capacity, continuous space, electromagnetic field,
information theory
\end{IEEEkeywords}

\section{Introduction}

Information theory has been widely applied to the analysis and design
of wireless communication systems to approach theoretical capacity
limits \cite{shannon1948mathematical}-\nocite{661517}\cite{telatar1999capacity}.
However, the physical realization of wireless communication systems
is based on the implementation of antennas and radio-frequency (RF)
circuits \cite{pozar2011microwave}, \cite{balanis2015antenna}. Therefore,
the joint study of information theory and electromagnetic field theory
can be utilized to further extend the theoretical channel capacity
limits of the entire wireless communication system. This has led to
the development of electromagnetic information theory (EIT) \cite{4636839},
\cite{4685903}. Utilizing EIT, the concept of aerial degrees-of-freedom
(ADoF) has been introduced to estimate the number of orthogonal basis
functions required for describing electromagnetic fields around antenna
systems \cite{bucci1989degrees}. The number of ADoF can be used to
determine the channel capacity because it refers to the maximum number
of parallel sub-channels that can be used for transmission \cite{poon2005degrees}.
To estimate the channel capacity bound for excitations restricted
to a given volume or area, continuous current sources limited to a
region, exciting the electromagnetic channel, have been analyzed using
continuous-space approaches \cite{migliore2006role}-\nocite{jensen2008capacity}\cite{jeon2017capacity}.
The results can be used to provide achievable bounds on the channel
capacity given a specific volume or area. However, they have not provided
the corresponding antenna designs to achieve those bounds.

Multiple-input multiple-output (MIMO) antenna systems, which exploit
ADoF by using multiple antennas at transceivers, play a critical role
in approaching the predicted channel capacity bounds in wireless communication
systems \cite{murch2002antenna}. As a result, an enormous number
of MIMO antenna designs have been proposed \cite{chiu2007compact}-\nocite{ren2014compact}\cite{zhai2015enhanced}.
A particular emphasis of these designs has been to devise approaches
that achieve as many antennas as possible within a certain volume
or area \cite{lu2014overview} and one example has provided up to
22 antennas per square wavelength \cite{soltani2015compact}. These
designs have to tradeoff strong mutual coupling effects with performance
and this has led to design limits on the antennas possible per unit
volume or area \cite{shen2015impedance}. As a result, the analysis
of the effects of mutual coupling on channel capacity performance
have also been well studied \cite{wallace2004mutual}, \cite{sun2011capacity}.
However there has been no direct link that can connect the design
of MIMO antennas with the continuous-space approaches utilized in
EIT \cite{4685903}, \cite{shyianov2021achievable}.

In this paper we introduce a method to link the antenna geometry and
continuous-space electromagnetic channel capacity together by using
loaded $N$-port structures \cite{harrington1972control}-\nocite{mautz1973modal}\cite{harrington1974pattern}.
It is an attempt to bridge the gap between the practical design of
MIMO antennas and their design limits predicted by EIT. This method
is not restricted to a specific antenna configuration and uses the
beamspace representation of MIMO systems. In beamspace MIMO, orthogonal
radiation patterns also form independent sub-channels in a similar
manner to the spatial separation of antennas that form spatial sub-channels
\cite{sayeed2002deconstructing}. Therefore, field distributions in
the far-field can be linked with the capacity analysis in beamspace
MIMO \cite{barousis2011beamspace}, \cite{barousis2011aerial} where
the closed-form expression for the channel capacity limits \cite{kalis2008novel},
\cite{alrabadi2009universal} are provided with various constraints.
Furthermore, by using loaded $N$-port structures to represent arbitrary
antenna geometries, an efficient optimization approach is proposed
for finding the optimum MIMO antenna design that achieves the channel
capacity bounds. Simulation results of the channel capacity bounds
and achieved capacity using our proposed antenna design with one wavelength
square size are provided. These show the designed antenna can achieve
capacity performance close to the fundamental bounds, demonstrating
the effectiveness of the proposed method.

\textit{Organization}: Section II formulates the MIMO system model
for the loaded $N$-port structures. Section III provides the derivation
for the resultant channel capacity with different constraints. Section
IV introduces the link with antenna design and describes efficient
optimization approaches to obtain the optimum antenna configuration.
In Section V, we provide numerical results of channel capacity for
a proposed MIMO antenna design to demonstrate the potential of the
technique. Section VI concludes the work.

\textit{Notation}: Bold lower and upper case letters denote vectors
and matrices respectively. Upper case letters in calligraphy denote
sets. Letters not in bold font represent scalars. $\mathrm{E}\left\{ a\right\} $
denotes the expectation of scalar $a$. $\left[\mathbf{a}\right]_{i}$
and $\left\Vert \mathbf{a}\right\Vert $ refer to the $i$th entry
and $l_{2}-$norm of vector $\mathrm{\mathbf{a}}$, respectively.
$\mathrm{\mathbf{A}}^{T}$, $\mathrm{\mathbf{A}}^{H}$, $\left[\mathrm{\mathbf{A}}\right]_{i,j}$
, $\left|\mathbf{A}\right|$ and $\mathrm{Tr}\left(\mathrm{\mathbf{A}}\right)$
refer to the transpose, conjugate transpose, $\left(i,j\right)$th
entry, determinant, and trace of a matrix $\mathrm{\mathbf{A}}$,
respectively. $\mathbb{C}$ denotes complex number sets and $j=\sqrt{-1}$
denotes an imaginary number. $\mathcal{CN}(\mu,\sigma^{2})$ denotes
complex Gaussian distribution with mean $\mu$ and variance $\sigma^{2}$.
$\mathbf{U}_{N}$ denotes an $N\times N$ identity matrix. diag$\left(a_{1},...,a_{N}\right)$
is a diagonal matrix with diagonal entries being $a_{1},...,a_{N}$.
$\left\langle \mathrm{\mathbf{a}},\mathrm{\mathbf{b}}\right\rangle =\mathrm{\mathbf{a}}^{H}\mathrm{\mathbf{b}}$
refers to the inner product of two vectors $\mathrm{\mathbf{a}}$
and $\mathrm{\mathbf{b}}$. It should also be noted that the dependence
of all variables on frequency is assumed and it is not explicitly
shown for brevity.

\section{System Model\label{sec:System-Model}}

\begin{figure}[t]
\begin{centering}
\textsf{\includegraphics[width=6cm]{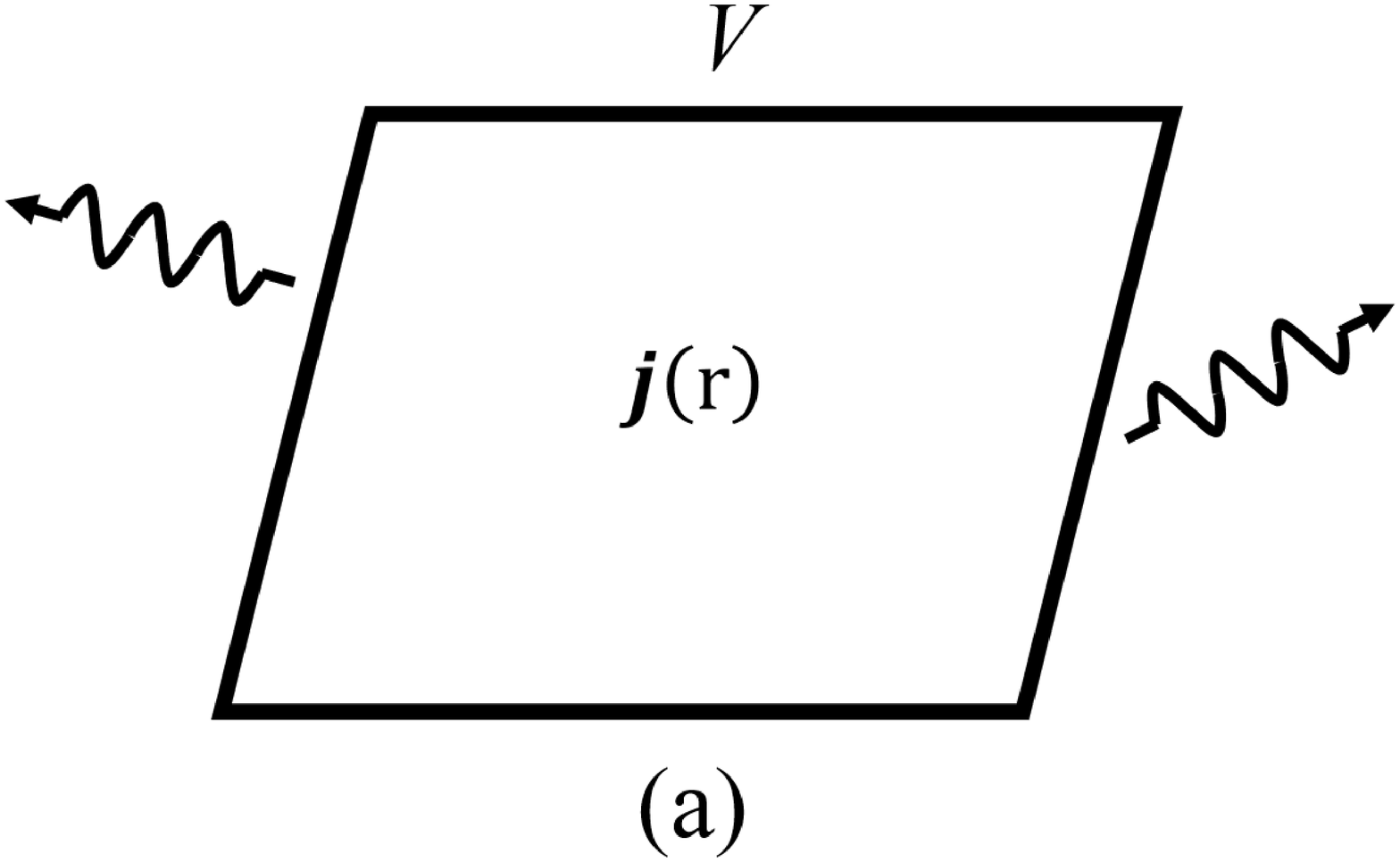}}
\par\end{centering}
\begin{centering}
\textsf{\includegraphics[width=6cm]{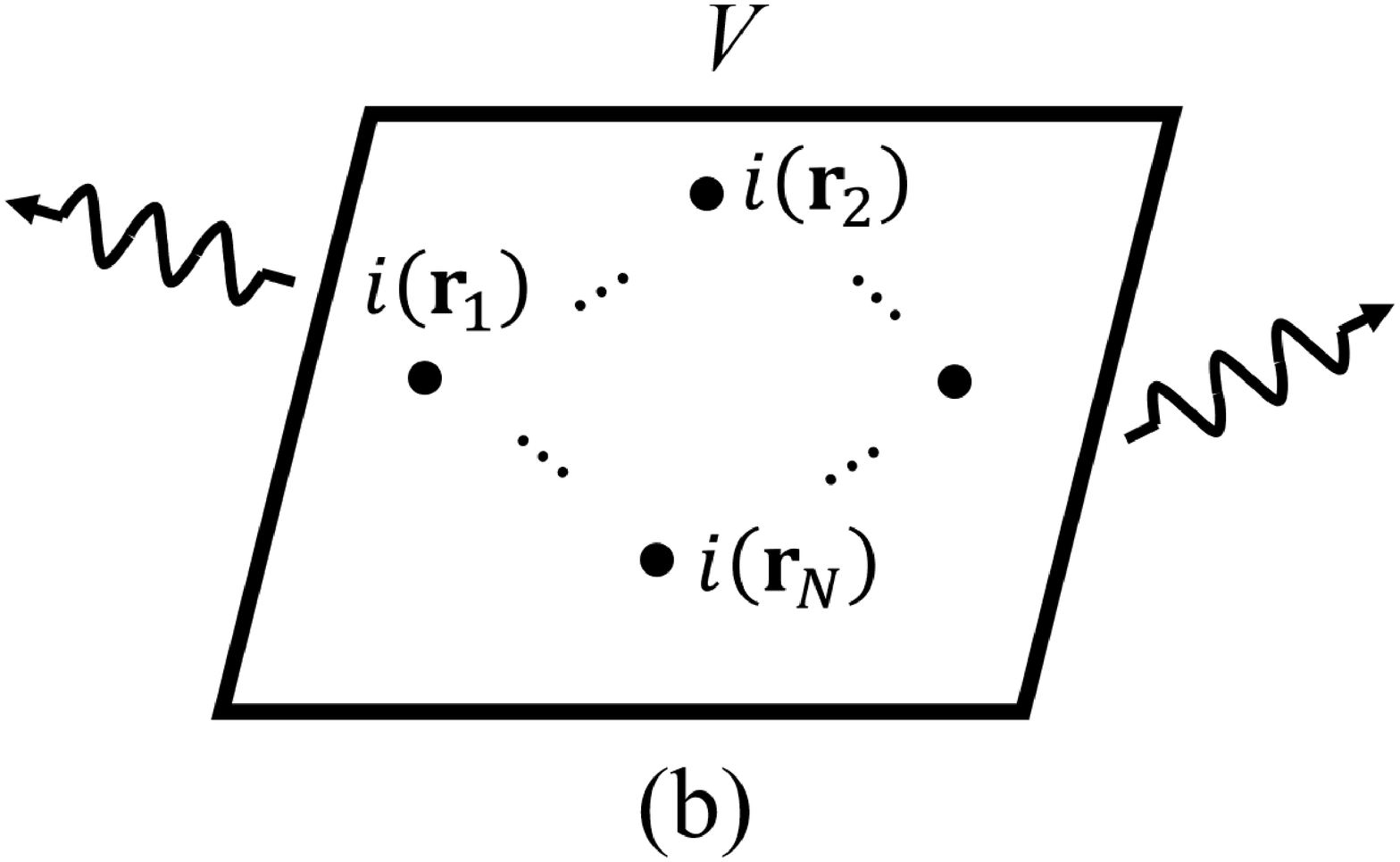}}
\par\end{centering}
\caption{Illustration of a current distribution $\mathbf{j}\left(\mathbf{r}\right)$
in a confined volume $V$ represented by (a) continuous space and
(b) approximated by a loaded $N$-port discretized structure.\label{fig:Illustrative pixel array}}
\end{figure}

In Fig. \ref{fig:Illustrative pixel array}(a) an arbitrary transmit
element is shown, confined to a volume $V$, with current distribution
$\mathbf{j}\left(\mathbf{r}\right)$, where $\mathbf{r}$ is the spatial
coordinate. When matched with the proper receive element and channel,
and with the current distribution $\mathbf{j}\left(\mathbf{r}\right)$
appropriately set, the resulting system can achieve the continuous-space
electromagnetic channel capacity bound \cite{jensen2008capacity}.
Our objective in this work is to design the optimum MIMO antenna that
can best approximate the required current distribution $\mathbf{j}\left(\mathbf{r}\right)$
in $V$ to approach the continuous-space electromagnetic channel capacity
bound. The challenge is to find structures that can create the necessary
current distribution $\mathbf{j}\left(\mathbf{r}\right)$ with the
discrete feeding ports required in MIMO antenna design.

Loaded structures are one approach that allow the formation of any
arbitrary current distribution on a surface \cite{harrington1972control}
and to connect this with the discrete feeds required in MIMO antenna
design, we can use $N$-port loaded structures \cite{mautz1973modal}
as also shown in Fig. \ref{fig:Illustrative pixel array}(b). In this
figure, the $N$ discrete ports are denoted by black dots \cite{mautz1973modal},
\cite{harrington1974pattern}. That is the volume $V$ is discretized
into ports each with currents $i\left(\mathbf{r}_{1}\right),i\left(\mathbf{r}_{2}\right),...,i\left(\mathbf{r}_{N}\right)$
\cite{harrington1974pattern}. By setting the separations between
ports in the structure to be significantly less than a wavelength,
it can approximate the required current distribution $\mathbf{j}\left(\mathbf{r}\right)$
by utilizing appropriate current excitations across the $N$ ports.
The advantage of this approach is that the feeds for the final potential
MIMO antenna are already incorporated at the beginning of the analysis
through the $N$-port loaded structure. In the final MIMO antenna
design, only some of the ports will be utilized as feeds with the
others loaded with reactances to form the required current distribution
that achieves the channel capacity bound. This allows us to form antennas
with the desired number of feeding ports while achieving performance
close to the channel capacity bounds.

For analysis we write $\boldsymbol{\mathrm{i}}=\left[i\left(\mathbf{r}_{1}\right),i\left(\mathbf{r}_{2}\right),...,i\left(\mathbf{r}_{N}\right)\right]{}^{T}\in\mathbb{C}^{N\times1}$
as the current at each of the ports in Fig. \ref{fig:Illustrative pixel array}(b).
Correspondingly the radiation patterns arising from each of the port
currents are written as $\mathbf{E}_{T}=\left[\mathbf{e}_{T,1},\mathbf{e}_{T,2},...,\mathbf{e}_{T,N}\right]\in\mathbb{C}^{K\times N}$
where $\mathbf{e}_{T,n}\in\mathbb{C}^{K\times1}$ is the radiation
pattern of the $n$th port excited by a unit current with all the
other ports open-circuited, over $K$ uniformly sampled three-dimensional
(3D) spatial angles. We use these patterns and the results in \cite{PCDM}
to decompose the radiation patterns into a natural set of orthonormal
modal functions similarly to the Theory of Characteristic Modes (TCM)
\cite{mautz1973modal}, \cite{9286866}. These orthonormal modal functions
can then be used for analysis of the channel capacity and interpreted
as being related to the ADoF of the structure.

For an arbitrary excitation current $\boldsymbol{\mathrm{i}}$, the
resulting radiation pattern can be expressed as $\mathbf{e}=\mathbf{E}_{T}\boldsymbol{\mathrm{i}}$,
so that the radiated power is given by
\begin{equation}
\frac{1}{\text{\ensuremath{\eta}}}\left\langle \mathbf{e},\mathbf{e}\right\rangle =\frac{1}{\text{\ensuremath{\eta}}}\left\langle \mathbf{E}_{T}\boldsymbol{\mathrm{i}},\mathbf{E}_{T}\boldsymbol{\mathrm{i}}\right\rangle =\frac{1}{\text{\ensuremath{\eta}}}\boldsymbol{\mathrm{i}}^{H}\mathbf{E}_{T}^{H}\mathbf{E}_{T}\boldsymbol{\mathrm{i}}=\boldsymbol{\mathrm{i}}^{H}\mathbf{K}_{T}\boldsymbol{\mathrm{i}},\label{eq:radiation power}
\end{equation}
where $\eta$ is the free space impedance, $\mathbf{K}_{T}$ is the
correlation of the steering matrix at the transmitter side \cite{PCDM},
which is defined as
\begin{equation}
\mathbf{K}_{T}=\frac{1}{\text{\ensuremath{\eta}}}\mathbf{E}_{T}^{H}\mathbf{E}_{T}.\label{eq:Kt}
\end{equation}
We can perform eigenvalue decomposition (EVD) on $\mathbf{K}_{T}$
as
\begin{equation}
\mathbf{K}_{T}=\mathbf{Q}\mathbf{\Lambda}\mathbf{Q}^{H},
\end{equation}
where $\Lambda=\mathrm{diag}\left(\lambda_{1},\lambda_{2},...,\lambda_{N}\right)$
is a real diagonal eigenvalue matrix. It is assumed that there is
no inner resonance in the transmit element \cite{mautz1973modal}
so that the loaded $N$-port structure can provide $N$ orthogonal
basis. Therefore, $\mathbf{K}_{T}$ is a positive definite matrix
with all eigenvalues being positive $\lambda_{1}\geqslant\lambda_{2}\geqslant...\geqslant\lambda_{N}>0$.
In practice some eigenvalues might be very small and this will affect
ADoF as discussed later in the paper. $\mathbf{Q}=\left[\mathbf{q}_{1},\mathbf{q}_{2},...,\mathbf{q}_{N}\right]$
is a unitary matrix whose column vectors $\mathbf{q}_{i}$, $i=1,2,...,N$,
are eigenvectors of $\mathbf{K}_{T}$ and form a set of current basis.
By selecting current $\boldsymbol{\mathrm{i}}$ to be $\mathbf{q}_{i}$
and collecting the corresponding radiation patterns $\mathbf{f}_{T,i}=\mathbf{E}_{T}\mathbf{q}_{i}$,
$i=1,2,...,N$, a set of orthogonal radiation patterns can be formed
since the inner product of two radiation patterns excited by $\mathbf{q}_{i}$
and $\mathbf{q}_{j}$ is given by
\begin{align}
\frac{1}{\text{\ensuremath{\eta}}}\left\langle \mathbf{f}_{T,i},\mathbf{f}_{T,j}\right\rangle  & =\frac{1}{\text{\ensuremath{\eta}}}\left\langle \mathbf{E}_{T}\mathbf{q}_{i},\mathbf{E}_{T}\mathbf{q}_{j}\right\rangle =\frac{1}{\text{\ensuremath{\eta}}}\mathbf{q}_{i}^{H}\mathbf{E}_{T}^{H}\mathbf{E}_{T}\mathbf{q}_{j}\nonumber \\
 & =\mathbf{q}_{i}^{H}\mathbf{Q}\mathbf{\Lambda}\mathbf{Q}^{H}\mathbf{q}_{j}=\delta_{ij}\lambda_{i},\label{eq:inner product}
\end{align}
where $\delta_{ij}$ is the Kronecker delta function (0 if $i\neq j$
and 1 if $i=j$). That is the radiation patterns excited by different
column vectors in $\mathbf{Q}$ are orthogonal. The $\lambda_{i}$,
$i=1,2,...,N$, can be regarded as the radiation resistance of the
$i$th radiation pattern since when excited by $\mathbf{q}_{i}$ (whose
$l_{2}-$norm is unity), the radiated power of $\mathbf{f}_{T,i}$
is $\lambda_{i}$.

The orthonormal basis set with unity radiated power can be written
as
\begin{equation}
\mathbf{B}_{T}=\mathbf{E}_{T}\mathbf{Q}\mathbf{\Lambda}^{-\frac{1}{2}}\label{eq:orthonormal basis}
\end{equation}
so that we have $\frac{1}{\text{\ensuremath{\eta}}}\mathbf{B}_{T}^{H}\mathbf{B}_{T}=\mathbf{U}_{N}$
with $\mathbf{B}_{T}=\left[\mathbf{b}_{T,1},\mathbf{b}_{T,2},...,\mathbf{b}_{T,N}\right]\in\mathbb{C}^{K\times N}$.
The diagonal entries in $\mathbf{\Lambda}^{-\frac{1}{2}}$ are scaling
factors to form the orthonormal basis. Therefore, for those basis
with small radiation resistances, large currents are required to radiate
patterns with unity power.

By replicating the approach for the receive volume, we respectively
denote the steering matrices of the $M$ receive antennas as $\mathbf{E}_{R}=\left[\mathbf{e}_{R,1},\mathbf{e}_{R,2},\ldots,\mathbf{e}_{R,M}\right]\in\mathbb{C}^{K\times M}$
where $\mathbf{e}_{R,m}\in\mathbb{C}^{K\times1}$, $m=1,2,...,M$.

The final step is to connect the transmitter and receiver using an
appropriate channel model. The beamspace domain or virtual channel
representation \cite{sayeed2002deconstructing}, \cite{maliatsos2013modifications}
is a well established approach that decomposes the channel into orthogonal
beams. This model fits naturally with the modal decompositions we
have formed at the transmitter and receiver. Using the beamspace approach,
the equivalent channel matrix in the angular domain can be expressed
as
\begin{equation}
\mathbf{H}=\frac{1}{\text{\ensuremath{\eta}}}\mathbf{E}_{R}^{H}\mathbf{H}_{v}\mathbf{E}_{T}\label{eq:angular channel}
\end{equation}
where $\mathbf{H}_{v}\in\mathbb{C}^{K\times K}$ is the virtual channel
whose entries refer to the channel gain from each angle of departure
(AoD) to each angle of arrival (AoA) \cite{maliatsos2013modifications}.
By taking \eqref{eq:angular channel} we can write the overall system
model as
\begin{align}
\mathbf{y} & =\frac{1}{\text{\ensuremath{\eta}}}\mathbf{E}_{R}^{H}\mathbf{H}_{v}\mathbf{E}_{T}\boldsymbol{\mathrm{i}}+\mathbf{n},\label{eq:angular MIMO system}
\end{align}
where $\mathbf{n}\in\mathbb{C}^{M\times1}$ is the additive Gaussian
noise and satisfies the complex Gaussian distribution $\mathcal{CN}\left(0,\sigma^{2}\mathbf{U}_{M}\right)$
with noise power $\sigma^{2}$ and $\mathbf{y}\in\mathbb{C}^{M\times1}$
is the received signal.

\section{Channel Capacity Formulation\label{sec:Channel-Capacity-Formulation}}

Using \eqref{eq:angular channel} and \eqref{eq:angular MIMO system},
channel capacity of the system in the beamspace domain can be written
as
\begin{align}
C & =\mathrm{log_{2}}\left|\mathbf{U}_{M}+\frac{\mathbf{H}\mathbf{R}_{\mathbf{i}}\mathbf{H}^{H}}{\sigma^{2}}\right|\label{eq:capacity}
\end{align}
where $\mathbf{R}_{\mathbf{i}}=\mathrm{E}\left\{ \mathbf{i}\mathbf{i}^{H}\right\} $
is the covariance matrix of current.

We assume the $M$ receive antennas ($N\leqslant M$ for tractability)
are ideally isolated so that the steering matrix $\mathbf{E}_{R}$
is an exactly orthonormal basis set of receive antennas satisfying
$\frac{1}{\text{\ensuremath{\eta}}}\mathbf{E}_{R}^{H}\mathbf{E}_{R}=\mathbf{U}_{M}$.
Using \eqref{eq:orthonormal basis}, we can rewrite the channel matrix
\eqref{eq:angular channel} as
\begin{equation}
\mathbf{H}=\frac{1}{\text{\ensuremath{\eta}}}\mathbf{E}_{R}^{H}\mathbf{H}_{v}\mathbf{B}_{T}\Lambda^{\frac{1}{2}}\mathbf{Q}^{H}=\mathbf{H}_{\mathrm{iid}}\Lambda^{\frac{1}{2}}\mathbf{Q}^{H},\label{eq:BS channel}
\end{equation}
where $\mathbf{H}_{\mathrm{iid}}=\frac{1}{\text{\ensuremath{\eta}}}\mathbf{E}_{R}^{H}\mathbf{H}_{v}\mathbf{B}_{T}\in\mathbb{C}^{M\times N}$
with each entry following i.i.d. distribution due to the orthonormality
among radiation patterns in $\mathbf{E}_{R}$ and $\mathbf{B}_{T}$.
The capacity \eqref{eq:capacity} is then rewritten using \eqref{eq:BS channel}
as
\begin{align}
C & =\mathrm{log_{2}}\left|\mathbf{U}_{M}+\frac{\mathbf{H}_{\mathrm{iid}}\Lambda^{\frac{1}{2}}\mathbf{Q}^{H}\mathbf{R}_{\mathbf{i}}\mathbf{Q}\Lambda^{\frac{1}{2}}\mathbf{H}_{\mathrm{iid}}^{H}}{\sigma^{2}}\right|.\label{eq:Capacity Simple 1}
\end{align}

Channel capacity is constrained by physical limits consisting of 1)
the radiated power constraint $\mathrm{Tr}\left(\frac{1}{\text{\ensuremath{\eta}}}\mathbf{E}_{T}\mathbf{R}_{\mathbf{i}}\mathbf{E}_{T}^{H}\right)=\mathrm{Tr}\left(\mathbf{R}_{\mathbf{i}}\mathbf{K}_{T}\right)\leqslant P_{\mathrm{rad}}$
with $P_{\mathrm{rad}}$ being the upper bound of radiated power and
2) the maximum currents that can exist at the input to the ports of
the reactively loaded structure, $\mathrm{Tr}\left(\mathbf{R}_{\mathbf{i}}\right)\leqslant I_{\mathrm{in}}^{2}$
with $I_{\mathrm{in}}$ being the upper bound of the current norm.
In practice, both constraints need to be applied to obtain an implementable
antenna. A constraint on the input power (it would be the same as
$P_{\mathrm{rad}}$ if the antenna was lossless) cannot be formulated
since an expression for the input impedances are not known as we do
not yet know the final feed arrangement.

In the following subsections, the formulation of capacity with two
constraints are derived, which is general and can be applied to arbitrary
$N$-port antenna systems.

\subsection{Capacity with Radiated Power Constraint\label{subsec:Pwr Cons Output}}

We firstly constrain the radiated power for the derivation of channel
capacity. This optimization problem can be formulated as
\begin{align}
\underset{\mathbf{R}_{\mathbf{i}}}{\mathrm{max}} & \quad\mathrm{log_{2}}\left|\mathbf{U}_{M}+\frac{\mathbf{H}_{\mathrm{iid}}\Lambda^{\frac{1}{2}}\mathbf{Q}^{H}\mathbf{R}_{\mathbf{i}}\mathbf{Q}\Lambda^{\frac{1}{2}}\mathbf{H}_{\mathrm{iid}}^{H}}{\sigma^{2}}\right|\label{eq:optimized capacity-2}\\
\mathrm{s.t.} & \quad\mathrm{Tr}\left(\mathbf{R}_{\mathbf{i}}\mathbf{K}_{T}\right)\leqslant P_{\mathrm{rad}}.\label{eq:optimized capacity constraint-2}
\end{align}

By decomposing arbitrary radiation pattern $\mathbf{e}$ onto the
set of orthonormal radiation pattern basis in $\mathbf{B}_{T}$, we
have
\begin{equation}
\mathbf{e}=\mathbf{E}_{T}\boldsymbol{\mathrm{i}}=\mathbf{B}_{T}\Lambda^{\frac{1}{2}}\mathbf{Q}^{H}\mathbf{i}=\mathbf{B}_{T}\boldsymbol{{\beta}}
\end{equation}
with
\begin{equation}
\boldsymbol{{\beta}}=\Lambda^{\frac{1}{2}}\mathbf{Q}^{H}\mathbf{i}\label{eq:Pr Cons Power Current Trans}
\end{equation}
so that each entry in $\boldsymbol{{\beta}}$ refers to the magnitude
allocated to each orthonormal pattern basis. The radiated power constraint
\eqref{eq:optimized capacity constraint-2} can be transformed to
\begin{align}
\mathrm{Tr}\left(\mathbf{R}_{\mathbf{i}}\mathbf{K}_{T}\right) & =\mathrm{Tr}\left(\Lambda^{\frac{1}{2}}\mathbf{Q}^{H}\mathbf{R}_{\mathbf{i}}\mathbf{Q}\Lambda^{\frac{1}{2}}\right)\nonumber \\
 & =\mathrm{Tr}\left(\mathrm{E}\left\{ \boldsymbol{{\beta}}\boldsymbol{{\beta}}^{H}\right\} \right)=\mathrm{Tr}\left(\mathbf{R}_{\beta}\right)\leqslant P_{\mathrm{rad}},\label{eq:Rbeta}
\end{align}
where $\mathbf{R}_{\beta}=\mathrm{E}\left\{ \boldsymbol{{\beta}}\boldsymbol{{\beta}}^{H}\right\} $
is the covariance matrix of orthonormal pattern basis magnitude. Accordingly,
the optimization problem \eqref{eq:optimized capacity-2} and \eqref{eq:optimized capacity constraint-2}
is re-formulated as
\begin{align}
\underset{\mathbf{R}_{\beta}}{\mathrm{max}} & \quad\mathrm{log_{2}}\left|\mathbf{U}_{M}+\frac{\mathbf{H}_{\mathrm{iid}}\mathbf{R}_{\beta}\mathbf{H}_{\mathrm{iid}}^{H}}{\sigma^{2}}\right|\label{eq:optimized capacity-2-1}\\
\mathrm{s.t.} & \quad\mathrm{Tr}\left(\mathbf{R}_{\beta}\right)\leqslant P_{\mathrm{rad}}.\label{eq:optimized capacity constraint-2-1}
\end{align}
It can be noticed that the problem \eqref{eq:optimized capacity-2-1}
and \eqref{eq:optimized capacity constraint-2-1} is the same as the
capacity maximization problem of ideal $M\times N$ MIMO. However,
due to some extremely small eigenvalues in $\Lambda$, the norm of
current $\mathbf{i}$ could be extremely large and not implementable.
In other words, the maximum current is unconstrained.

\subsection{Capacity with Current Constraint\label{subsec:Pwr Cons Input}}

Next, we consider imposing a constraint on the current norm $\mathrm{Tr}\left(\mathbf{R}_{\mathbf{i}}\right)$.
The optimization problem of channel capacity can then be formulated
as
\begin{align}
\underset{\mathbf{R}_{\mathbf{i}}}{\mathrm{max}} & \quad\mathrm{log_{2}}\left|\mathbf{U}_{M}+\frac{\mathbf{H}_{\mathrm{iid}}\Lambda^{\frac{1}{2}}\mathbf{Q}^{H}\mathbf{R}_{\mathbf{i}}\mathbf{Q}\Lambda^{\frac{1}{2}}\mathbf{H}_{\mathrm{iid}}^{H}}{\sigma^{2}}\right|\label{eq:optimized capacity-1}\\
\mathrm{s.t.} & \quad\mathrm{Tr}\left(\mathbf{R}_{\mathbf{i}}\right)\leqslant I_{\mathrm{in}}^{2}.\label{eq:optimized capacity constraint-1}
\end{align}

By decomposing the current $\mathbf{i}$ onto the set of current basis
$\mathbf{q}_{i}$, $i=1,2,...,N$, we can obtain the coefficient for
the decomposition as $\gamma_{i}=\left\langle \mathbf{q}_{i},\mathbf{i}\right\rangle =\mathbf{q}_{i}^{H}\mathbf{i}$
which refers to the magnitude allocated to the $i$th current basis.
We collect the coefficient $\gamma_{i}$ into a vector $\boldsymbol{{\gamma}}=\left[\gamma_{1},\gamma_{2},...,\gamma_{N}\right]^{T}$
which can be related to $\mathbf{i}$ by
\begin{equation}
\boldsymbol{{\gamma}}=\mathbf{Q}^{H}\mathbf{i}.\label{eq:Pr Cons Current Current Trans}
\end{equation}
$\boldsymbol{{\gamma}}$ has the same norm as $\mathbf{i}$ so that
the current constraint \eqref{eq:optimized capacity constraint-1}
can be transformed to 
\begin{align}
\mathrm{Tr}\left(\mathbf{R}_{\mathbf{i}}\right) & =\mathrm{Tr}\left(\mathbf{Q}^{H}\mathbf{R}_{\mathbf{i}}\mathbf{Q}\right)\nonumber \\
 & =\mathrm{Tr}\left(\mathrm{E}\left\{ \boldsymbol{{\gamma}}\boldsymbol{{\gamma}}^{H}\right\} \right)=\mathrm{Tr}\left(\mathbf{R}_{\gamma}\right)\leqslant I_{\mathrm{in}}^{2}\text{,}\label{eq:R alpha}
\end{align}
where $\mathbf{R}_{\gamma}=\mathrm{E}\left\{ \boldsymbol{{\gamma}}\boldsymbol{{\gamma}}^{H}\right\} $
is the covariance matrix of current basis magnitude. Then the problem
\eqref{eq:optimized capacity-1} and \eqref{eq:optimized capacity constraint-1}
can be transformed to
\begin{align}
\underset{\mathbf{R}_{\gamma}}{\mathrm{max}} & \quad\mathrm{log_{2}}\left|\mathbf{U}_{M}+\frac{\mathbf{H}_{\mathrm{iid}}\Lambda^{\frac{1}{2}}\mathbf{R}_{\gamma}\Lambda^{\frac{1}{2}}\mathbf{H}_{\mathrm{iid}}^{H}}{\sigma^{2}}\right|\label{eq:optimized capacity-1-1}\\
\mathrm{s.t.} & \quad\mathrm{Tr}\left(\mathbf{R}_{\gamma}\right)\leqslant I_{\mathrm{in}}^{2}.\label{eq:optimized capacity constraint-1-1}
\end{align}
The optimal solution for the currents can be found by performing equal
power (EP) or water-filling (WF) allocation directly on $\mathbf{R}_{\gamma}$.

It can be observed that due to the existence of $\Lambda^{\frac{1}{2}}$
in \eqref{eq:optimized capacity-1-1}, the radiated power of each
orthogonal basis in $\mathbf{B}_{T}$ is not equal and proportional
to the corresponding eigenvalues in $\Lambda$. Therefore, in this
formulation, the WF method tends to allocate more power to those basis
with larger eigenvalues since it increases radiated power and resulting
system capacity. In other words, the radiated power is unconstrained.

\subsection{Capacity with Dual Constraint\label{subsec:Capacity-with-Dual}}

Finally, we consider dual constraints imposed by the current norm
and radiated power simultaneously. That is, in practical setups the
radiated power is limited and the norm of current $\mathbf{i}$ must
also be restricted. This optimization problem is formulated as
\begin{align}
\underset{\mathbf{R}_{\mathbf{i}}}{\mathrm{max}} & \quad\mathrm{log_{2}}\left|\mathbf{U}_{M}+\frac{\mathbf{H}_{\mathrm{iid}}\Lambda^{\frac{1}{2}}\mathbf{Q}^{H}\mathbf{R}_{\mathbf{i}}\mathbf{Q}\Lambda^{\frac{1}{2}}\mathbf{H}_{\mathrm{iid}}^{H}}{\sigma^{2}}\right|\label{eq:optimized capacity-3}\\
\mathrm{s.t.} & \quad\mathrm{Tr}\left(\mathbf{R}_{\mathbf{i}}\mathbf{K}_{T}\right)\leqslant P_{\mathrm{rad}},\label{eq:optimized capacity constraint-3-1}\\
 & \quad\mathrm{Tr}\left(\mathbf{R}_{\mathbf{i}}\right)\leqslant I_{\mathrm{in}}^{2}.\label{eq:optimized capacity constraint-3-2}
\end{align}

We follow the transformation in \eqref{eq:Pr Cons Power Current Trans}
as well as the formulation in \eqref{eq:optimized capacity-2-1} and
\eqref{eq:optimized capacity constraint-2-1}, so the problem \eqref{eq:optimized capacity-3}
to \eqref{eq:optimized capacity constraint-3-2} can be transformed
to
\begin{align}
\underset{\mathbf{R}_{\beta}}{\mathrm{max}} & \quad\mathrm{log_{2}}\left|\mathbf{U}_{M}+\frac{\mathbf{H}_{\mathrm{iid}}\mathbf{R}_{\beta}\mathbf{H}_{\mathrm{iid}}^{H}}{\sigma^{2}}\right|\label{eq:optimized capacity-3-1}\\
\mathrm{s.t.} & \quad\mathrm{Tr}\left(\mathbf{R}_{\beta}\mathbf{\Lambda}^{-1}\right)\leqslant I_{\mathrm{in}}^{2},\label{eq:optimized capacity constraint-3-1-1}\\
 & \quad\mathrm{Tr}\left(\mathbf{R}_{\beta}\right)\leqslant P_{\mathrm{rad}}.\label{eq:optimized capacity constraint-3-2-1}
\end{align}
It can be observed that the capacity formulation is only related to
$\Lambda$, which is a diagonal matrix with diagonal entries being
the eigenvalues of $\mathbf{K}_{T}$.

To solve the problem \eqref{eq:optimized capacity-3-1} to \eqref{eq:optimized capacity constraint-3-2-1},
we firstly consider EP allocation method. Although the system provides
$N$ orthogonal radiation pattern basis, some eigenvalues in $\Lambda$
are too small so that large currents are required to radiate their
corresponding basis \eqref{eq:orthonormal basis} which cannot be
used for practical MIMO antenna design. To avoid generating large
currents, i.e. satisfying the current constraint \eqref{eq:optimized capacity constraint-3-2-1},
$P_{\mathrm{rad}}$ should be allocated to those basis with largest
radiation resistance (i.e. larger eigenvalues in $\Lambda$). Therefore,
we equally allocate power $P_{\mathrm{rad}}$ to the first $N_{\mathrm{eff}}$
$\left(N_{\mathrm{eff}}\leqslant N\right)$ basis and drop the remaining
$N-N_{\mathrm{eff}}$ basis. These $N_{\mathrm{eff}}$ basis can be
effectively used with maximum current norm $I_{\mathrm{in}}^{2}$
and radiated power $P_{\mathrm{rad}}$, and thus can be regarded as
the effective aerial degrees-of-freedom (EADoF) of the transmitter
\cite{barousis2011aerial}.

The diagonal entries in $\mathbf{R}_{\beta}$ are then given by
\begin{equation}
\left[\mathbf{R}_{\beta}\right]_{i,i}=\begin{cases}
\frac{P_{\mathrm{rad}}}{N_{\mathrm{eff}}}, & i=1,2,...,N_{\mathrm{eff}},\\
0, & \mathrm{otherwise.}
\end{cases}
\end{equation}
The constraint \eqref{eq:optimized capacity constraint-3-2-1} can
then be written as
\begin{equation}
\mathrm{Tr}\left(\mathbf{R}_{\beta}\Lambda^{-1}\right)=\stackrel[i=1]{N_{\mathrm{eff}}}{\sum}\frac{P_{\mathrm{rad}}}{N_{\mathrm{eff}}}\mathbf{\lambda}_{i}^{-1}\leqslant I_{\mathrm{in}}^{2}.\label{eq:optimized capacity constraint-3-1-1-1}
\end{equation}
We define $\epsilon=\frac{I_{\mathrm{in}}^{2}}{P_{\mathrm{rad}}}$
in \eqref{eq:optimized capacity constraint-3-1-1-1} and it can be
interpreted loosely as a conductance which has the unit of $\Omega^{-1}$.
In essence for a given $P_{\mathrm{rad}}$, a larger $I_{\mathrm{in}}$
($\epsilon$ high) indicates that the antenna will have a smaller
input resistance overall and if it is too small the antenna will not
be implementable. Alternatively if $I_{\mathrm{in}}$ is too low ($\epsilon$
low), the input impedance will be required to be too high and also
not implementable. Therefore we should set it to be in a range centered
around 1/50 $\Omega^{-1}$ so that it is in line with the antennas
desired input impedance.

Then the EADoF $N_{\mathrm{eff}}$ is related to $\epsilon$ for a
fixed SNR as
\begin{equation}
\stackrel[i=1]{N_{\mathrm{eff}}}{\sum}\frac{\mathbf{\lambda}_{i}^{-1}}{N_{\mathrm{eff}}}\leqslant\epsilon\label{eq:EP for Dual Cons}
\end{equation}
where the left term refers to the average of the reciprocal of eigenvalues
for the first $N_{\mathrm{eff}}$ basis. It can be observed from \eqref{eq:EP for Dual Cons}
that a larger $\epsilon$ indicates that the number of available orthogonal
basis, i.e. $N_{\mathrm{eff}}$, and the consequent capacity bound
can be larger. The maximum $N_{\mathrm{eff}}$ can be obtained by
solving \eqref{eq:EP for Dual Cons} for a fixed $\epsilon$. That
is EADoF $N_{\mathrm{eff}}$ is dependent on $\epsilon$ (the maximum
allowable current norm given $P_{\mathrm{rad}}$), indicating the
number of basis that can be used with the constraint of current norm
and radiation power.

Next we consider using the WF method to maximize capacity in problem
\eqref{eq:optimized capacity-3-1} to \eqref{eq:optimized capacity constraint-3-2-1}.
We start with the channel gain of $N$ sub-channels in channel matrix
$\mathbf{H}_{\mathrm{iid}}$. By performing EVD on $\mathbf{H}_{\mathrm{iid}}\mathbf{H}_{\mathrm{iid}}^{H}$,
we obtain the channel gain matrix $\mathbf{S}=\mathrm{diag}\left(s_{1},s_{2},...,s_{N}\right)$
which is a real diagonal matrix with $s_{i}$, $i=1,2,...,N$, being
the channel gain of the $i$th sub-channel. Then we use the Lagrangian
method whose function is given by
\begin{align}
L= & -\mathrm{log_{2}}\left|\mathbf{U}_{N}+\frac{\mathbf{S}}{\sigma^{2}}\mathbf{R}_{\beta}\right|+\mu_{\mathrm{rad}}\mathrm{Tr}\left(\mathbf{R}_{\beta}-\frac{P_{\mathrm{rad}}}{N}\mathbf{U}_{N}\right)\nonumber \\
 & +\mu_{\mathrm{in}}\mathrm{Tr}\left(\mathbf{R}_{\beta}\Lambda^{-1}-\frac{I_{\mathrm{in}}^{2}}{N}\mathbf{U}_{N}\right),\label{eq:Lagrangian}
\end{align}
where $\mu_{\mathrm{rad}}$ and $\mu_{\mathrm{in}}$ are Lagrangian
multipliers. Taking the partial derivative of $L$ with respect to
$\mathbf{R}_{\beta}$, we have
\begin{align}
-\mathrm{log_{2}}e\cdot\mathbf{S}\left(\sigma^{2}\mathbf{U}_{N}+\mathbf{S}\mathbf{R}_{\beta}\right)^{-1} & +\mu_{\mathrm{rad}}\mathbf{U}_{N}+\mu_{\mathrm{in}}\Lambda^{-1}=\mathbf{0}.\label{eq:Partial Derivative}
\end{align}
Combining the solution of $\mathbf{R}_{\beta}$ in \eqref{eq:Partial Derivative},
two constraints \eqref{eq:optimized capacity constraint-3-1-1}, \eqref{eq:optimized capacity constraint-3-2-1}
can be written as \cite{sun2011capacity}
\begin{equation}
\mathrm{Tr}\left(\mathrm{log_{2}}e\left(\mu_{\mathrm{rad}}\mathbf{U}_{N}+\mu_{\mathrm{in}}\Lambda^{-1}\right)^{-1}-\sigma^{2}\mathbf{S}^{-1}\right)=P_{\mathrm{rad}},\label{eq:Constraint WF C1}
\end{equation}
\begin{align}
\mathrm{Tr}\left(\mathrm{log_{2}}e\left(\mu_{\mathrm{rad}}\Lambda+\mu_{\mathrm{in}}\mathbf{U}_{N}\right)^{-1}-\sigma^{2}\left(\Lambda\mathbf{S}\right)^{-1}\right) & =I_{\mathrm{in}}^{2}.\label{eq:Constraint WF C2}
\end{align}
The optimal multipliers $\mu_{\mathrm{rad}}^{\star}$ and $\mu_{\mathrm{in}}^{\star}$
in \eqref{eq:Constraint WF C2} and \eqref{eq:Constraint WF C1} can
be found via binary search \cite{sun2011capacity} and the optimal
$\mathbf{R}_{\beta}$ can be solved in \eqref{eq:Partial Derivative}.

It should be noted that the above closed-form expressions for system
capacity are formulated without considering the physical realization
and practical excitation for MIMO antennas. Therefore, the MIMO antenna
structure should be carefully designed and optimized to achieve the
capacity bounds with dual constraints, which will be described in
the next section.

\section{Loaded $N$-port Structure and Analysis \label{sec:Antenna Design}}

In the previous section, we have provided an optimization formulation
to obtain the required port currents on the loaded $N$-port structure
for capacity maximization with various constraints. In this section
we link those currents to MIMO antenna design with only $Q<<N$ feeding
ports in the loaded $N$-port structures.

\subsection{Network Analysis}

\begin{figure}[t]
\begin{centering}
\textsf{\includegraphics[width=8.5cm]{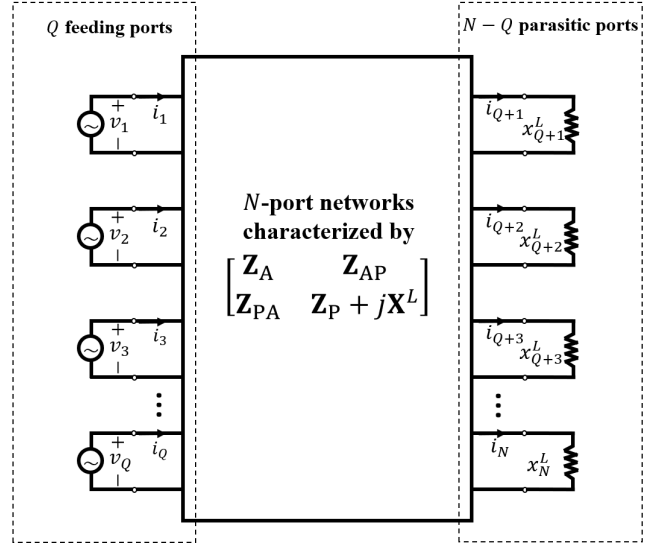}}
\par\end{centering}
\caption{Equivalent circuit model of the $N$-port network with $Q$ feeding
ports and $N-Q$ parasitic ports.\label{fig:Equivalent circuit-1}}
\end{figure}

To begin to construct the MIMO antenna, we divide the $N$-ports in
Fig. \ref{fig:Illustrative pixel array}(b) into $Q$ active feeding
ports and $N-Q$ loaded ports. The advantage of this approach is that
the feeds for the final potential $Q$-port MIMO antenna are incorporated
at the beginning of the analysis through the $N$-port structure.
The equivalent circuit model for the $N$-port network with $Q$ active
feeding ports (numbered 1 to $Q$) and $N-Q$ parasitic loaded ports
(numbered $Q+1$ to $N$) is shown in Fig. \ref{fig:Equivalent circuit-1}.
The loaded ports are either lossless with inductive and capacitive
reactance or open. The active feeding ports are connected to a matching
network to achieve the necessary 50 $\Omega$ input impedance. By
optimizing the $N-Q$ load reactances, we wish to generate $Q$ orthogonal
radiation patterns from the designated $Q$ feeding ports so that
the capacity \eqref{eq:capacity} can be maximized. In essence we
will attempt to find loads that create the required current distributions
for approaching the capacity bounds. The design procedure for selecting
which ports are fed and which are loaded to achieve the optimum channel
capacity using \eqref{eq:Capacity Simple 1} is described later.

In Fig. \ref{fig:Equivalent circuit-1}, the $q$th feeding port is
excited by a voltage source $v_{q}$, $q=1,2,...,Q$, and the $\left(Q+p\right)$th
parasitic port is loaded with reactance $x_{Q+p}^{L}$, $p=1,2,...,N-Q$.
We group the voltage source excitation $v_{q}$, $q=1,2,...,Q$, into
a vector as $\mathbf{v}_{\mathrm{A}}=\left[v_{1},v_{2},...,v_{Q}\right]{}^{T}\in\mathbb{C}^{Q\times1}$
and define $\mathbf{v}_{\mathrm{P}}=\left[0,0,...,0\right]{}^{T}\in\mathbb{C}^{\left(N-Q\right)\times1}$
as a $\left(N-Q\right)$-dimension zero vector due to there being
no excitation at the parasitic ports. We also group current at the
feeding and parasitic ports into vectors as $\mathbf{i}_{\mathrm{A}}=\left[i_{1},i_{2},...,i_{Q}\right]{}^{T}\in\mathbb{C}^{Q\times1}$
and $\mathbf{i}_{\mathrm{P}}=\left[i_{Q+1},i_{Q+2},...,i_{N}\right]{}^{T}\in\mathbb{C}^{\left(N-Q\right)\times1}$,
respectively. The voltage and current in the $N$-port network are
then related by
\begin{equation}
\left[\begin{array}{c}
\mathbf{v}_{\mathrm{A}}\\
\mathbf{v}_{\mathrm{P}}
\end{array}\right]=\left[\begin{array}{ll}
\mathbf{Z}_{\mathrm{A}} & \mathbf{Z}_{\mathrm{AP}}\\
\mathbf{Z}_{\mathrm{PA}} & \mathbf{Z}_{\mathrm{P}}+j\mathrm{\mathbf{X}}^{L}
\end{array}\right]\left[\begin{array}{c}
\mathbf{i}_{\mathrm{A}}\\
\mathbf{i}_{\mathrm{P}}
\end{array}\right],\label{eq:V=00003DZI}
\end{equation}
where $\mathrm{\mathbf{X}}^{L}=\mathrm{diag}\mathnormal{\left(x_{Q+\text{1}}^{L},x_{Q+\text{2}}^{L},\cdots,x_{N}^{L}\right)}$
represents the load reactance connected to each parasitic port. $\mathbf{Z}_{\mathrm{A}}\in\mathbb{C}^{Q\times Q}$
and $\mathbf{Z}_{\mathrm{P}}\in\mathbb{C}^{\left(N-Q\right)\times\left(N-Q\right)}$
are the impedance sub-matrix of the $Q$ feeding ports and the $N-Q$
parasitic ports, respectively. $\mathbf{Z}_{\mathrm{AP}}\in\mathbb{C}^{Q\times\left(N-Q\right)}$
and $\mathbf{Z}_{\mathrm{PA}}\in\mathbb{C}^{\left(N-Q\right)\times Q}$
are the sub-matrix referring to the mutual impedance between $Q$
feeding and $N-Q$ parasitic ports with $\mathbf{Z}_{\mathrm{AP}}=\mathbf{Z}_{\mathrm{PA}}^{T}$.
From \eqref{eq:V=00003DZI} we can obtain the relationship between
the current on the feeding and parasitic ports as
\begin{equation}
\mathbf{i}_{\mathrm{P}}=-\left(\mathbf{Z}_{\mathrm{P}}+j\mathrm{\mathbf{X}}^{L}\right)^{-1}\mathbf{Z}_{\mathrm{PA}}\mathbf{i}_{\mathrm{A}}.\label{eq:current relation}
\end{equation}

Our interest is obtaining $Q$ orthogonal far-field radiation patterns
to form the beamspace MIMO system. To that end, we collect the radiation
pattern of each feeding port $\mathrm{\mathbf{e}}_{T,q}\left(\Omega\right)$,
$q=1,2,...,Q$, in the antenna system into matrix form $\mathrm{\mathbf{E}}_{T}=\left[\mathrm{\mathbf{e}}_{T,1}\left(\Omega\right),\mathrm{\mathbf{e}}_{T,2}\left(\Omega\right),...,\mathrm{\mathbf{e}}_{T,Q}\left(\Omega\right)\right]$,
which can be given as
\begin{equation}
\mathrm{\mathbf{E}}_{T}\left(\Omega,\mathrm{\mathbf{X}}^{L}\right)=\mathbf{E}_{\mathrm{A}}\left(\Omega\right)-\mathbf{E}_{\mathrm{P}}\left(\Omega\right)\left(\mathbf{Z}_{\mathrm{P}}+j\mathrm{\mathbf{X}}^{L}\right)^{-1}\mathbf{Z}_{\mathrm{PA}},\label{eq:F pattern}
\end{equation}
where $\Omega=\left(\theta,\phi\right)$ denotes the spatial angle
with $\theta$ and $\phi$ representing the elevation and azimuth
angles in spherical coordinates, respectively. $\mathbf{E}_{\mathrm{A}}\left(\Omega\right)=\left[\mathbf{e}_{\mathrm{A},1}\left(\Omega\right),\ldots,\mathbf{e}_{\mathrm{A},Q}\left(\Omega\right)\right]$
collects $Q$ open-circuit radiation patterns of the feeding ports
with $\mathbf{e}_{\mathrm{A},q}\left(\Omega\right)$, $q=1,2,...,Q$
being the open-circuit radiation pattern of the $q$th feeding port
excited by a unit current when all the other feeding and parasitic
ports are open-circuit. Similarly, $\mathbf{E}_{\mathrm{P}}\left(\Omega\right)=\left[\mathbf{e}_{\mathrm{P},Q+1}\left(\Omega\right),\ldots,\mathbf{e}_{\mathrm{P},N}\left(\Omega\right)\right]$
collects $N-Q$ open-circuit radiation patterns of the parasitic ports
with $\mathbf{e}_{\mathrm{P},Q+p}\left(\Omega\right)$, $p=1,2,...,N-Q$,
being the open-circuit radiation pattern of the $\left(Q+p\right)$th
parasitic port excited by a unit current when all the other feeding
and parasitic ports are open-circuited. It can be observed that $\mathrm{\mathbf{E}}_{T}$
consists of the original open-circuit radiation pattern of the $Q$
feeding ports $\mathbf{E}_{\mathrm{A}}$ and a perturbation term affected
by the load reactance $\mathrm{\mathbf{X}}^{L}$ across the parasitic
ports.

We use a full electromagnetic solver, CST studio suite \cite{CST},
to simulate the open-circuit radiation patterns of all $N$ ports
of the loaded $N$-port structure. It should be noted that the simulation
only needs to be performed once because any radiation pattern excited
by any current distribution can then be found by using \eqref{eq:F pattern}
\cite{zhang2021DGS}. This reduces the computation complexity enormously
as full electromagnetic simulation is not needed during the load optimization
and capacity simulation stages.

\subsection{Optimization}

To obtain the optimal design of the MIMO antenna, we wish to select
$Q$ active feeding ports and optimize the load reactances at the
$N-Q$ parasitic ports. The objective is to generate a set of orthogonal
radiation patterns in $\mathrm{\mathbf{E}}_{T}$ at $Q$ feeding ports
so that the resulting channel capacity can be maximized.

To achieve an optimization result that also meets practical design
constraints, we need to note that not all ports would be suitable
for use as feeding. For example, ports positioned on the edges of
the structure are more likely to be accessible for feeds than those
completely surrounded by ports. For this reason, we partition the
$N$ ports into two sets for optimization. Assume $S$ $\left(S>Q\right)$
out of $N$ ports are feasible feeding ports and whose indices in
the set are given by $\mathcal{S}=\left\{ 1,2,...,S\right\} $. The
objective is therefore to find the optimum $Q$ ports from the $S$
feasible ports where the set of indices for the $Q$ feeding ports
is given by $\mathcal{P}=\left\{ p_{1},p_{2},...,p_{Q}\right\} \subset\mathcal{S}$
. In addition, we must find the optimum loads for the remaining $N-S$
ports so that together with the $S-Q$ unselected feeding ports, the
load reactance matrix $\mathbf{X}^{L}$ can meet the requirements
of orthogonality for the $Q$ selected feeding ports.

We use the correlation coefficient between the radiation patterns
of two feeding ports, i.e. $\text{\ensuremath{\mathrm{\mathbf{e}}_{T,j}}}\left(\Omega,\mathrm{\mathbf{X}}^{L}\right)$
of the $j$th feeding port and $\text{\ensuremath{\mathrm{\mathbf{e}}_{T,k}}}\left(\Omega,\mathrm{\mathbf{X}}^{L}\right)$
of the $k$th feeding port, to evaluate their similarity, and this
is defined as
\begin{equation}
\rho_{jk}\left(\mathrm{\mathbf{X}}^{L}\right)=\frac{\left\langle \text{\ensuremath{\mathrm{\mathbf{e}}_{T,j}}}\left(\Omega,\mathrm{\mathbf{X}}^{L}\right),\text{\ensuremath{\mathrm{\mathbf{e}}_{T,k}}}\left(\Omega,\mathrm{\mathbf{X}}^{L}\right)\right\rangle }{\left\Vert \mathrm{\mathbf{e}}_{T,j}\left(\Omega,\mathrm{\mathbf{X}}^{L}\right)\right\Vert \left\Vert \mathrm{\mathbf{e}}_{T,k}\left(\Omega,\mathrm{\mathbf{X}}^{L}\right)\right\Vert }\label{eq:CorrCoef}
\end{equation}
where $\rho_{jk}$ satisfies $0\leq\left|\rho_{jk}\right|\leq1$.
When $\left|\rho_{jk}\right|=0$, $\text{\ensuremath{\mathrm{\mathbf{e}}_{T,j}}}$
and $\text{\ensuremath{\mathrm{\mathbf{e}}_{T,k}}}$ are orthogonal
to each other. Leveraging the correlation coefficient \eqref{eq:CorrCoef},
we can formulate the optimization problem as
\begin{align}
\underset{\mathcal{P},\:\mathbf{X}^{L}}{\mathrm{min}}\:\:\:\: & \stackrel[k\in\mathcal{P}]{j\neq k}{\sum}\stackrel[j\in\mathcal{P}]{}{\sum}\left|\rho_{jk}\left(\mathrm{\mathbf{X}}^{L}\right)\right|.\label{eq:optimum ports with loads}
\end{align}
It should be noted that the optimization variables $\mathcal{P}$
and $\mathbf{X}^{L}$ are highly coupled with each other in this problem.
This is because altering feeding port indices greatly changes the
impedance matrix $\mathbf{Z}_{\mathrm{A}}$, $\mathbf{Z}_{\mathrm{AP}}$
and $\mathbf{Z}_{\mathrm{P}}$ in \eqref{eq:V=00003DZI}, making the
problem \eqref{eq:optimum ports with loads} complicated to solve.
To overcome this difficulty and meet these constraints, we propose
the use of the alternating optimization method to iteratively optimize
the load reactances and selection of the feeding ports \cite{bingbing2014indoor},
\cite{bezdek2003convergence}. For simplicity, the $S-Q$ unselected
feeding ports are left open in the entire optimization process to
ease the computational burden. However they could also be included
in the load optimization process if deemed necessary.

To start with, we randomly select $Q$ feeding ports, out of $S$,
as the initial guess at iteration 0. The set consisting of the indices
for $Q$ initial random feeding ports is given by $\mathcal{P}^{\left(0\right)}=\left\{ p_{1}^{\left(0\right)},p_{2}^{\left(0\right)},...,p_{Q}^{\text{\ensuremath{\left(0\right)}}}\right\} \subset\mathcal{S}$
with $p_{q}^{\left(0\right)}$ being the index of the $q$th initial
feeding port. Presetting the $S-Q$ unused feeding ports to be open,
we then need to optimize $N-S$ load reactances at the parasitic ports
where the initial guess for the load reactance matrix is given as
$\mathbf{X}^{L\left(0\right)}=\mathrm{diag}\mathnormal{\left(\infty,...,\infty,\text{0},...,\text{0}\right)}$
with the first $S-Q$ diagonal entries being fixed as infinity (open).

At the $i$th iteration, we wish to find the load reactances $\mathbf{X}^{L\left(i\right)}$
to produce near orthogonal radiation patterns with feeding ports indices
set $\mathcal{P}^{\left(i-1\right)}$. This can be performed by solving
the optimization problem
\begin{align}
\underset{\mathbf{X}^{L\left(i\right)}}{\mathrm{min}}\:\:\:\: & \stackrel[k\in\mathcal{P}^{\left(i-1\right)}]{j\neq k}{\sum}\stackrel[j\in\mathcal{P}^{\left(i-1\right)}]{}{\sum}\left|\rho_{jk}\left(\mathrm{\mathbf{X}}^{L\left(i\right)}\right)\right|.\label{eq:optimum}
\end{align}
This makes the correlation coefficient norm between any two patterns
in $\mathrm{\mathbf{E}}_{T}$ as close to zero as possible. To solve
the unconstrained optimization problem \eqref{eq:optimum}, we can
use the quasi-Newton method \cite{gill1972quasi} which guarantees
convergence to a stationary point of the problem \eqref{eq:optimum}.
The optimal load reactances at the $i$th iteration are denoted by
$\mathbf{X}^{L\left(i\right)}$.

We then use $\mathbf{X}^{L\left(i\right)}$ from \eqref{eq:optimum}
to optimize the indices of the feeding ports $\mathcal{P}^{\left(i\right)}$,
that is, 
\begin{align}
\underset{\mathcal{P}^{\left(i\right)}}{\mathrm{min}}\:\:\:\: & \stackrel[k\in\mathcal{P}^{\left(i\right)}]{j\neq k}{\sum}\stackrel[j\in\mathcal{P}^{\left(i\right)}]{}{\sum}\left|\rho_{jk}\left(\mathrm{\mathbf{X}}^{L\left(i\right)}\right)\right|,\label{eq:optimum ports}
\end{align}
which is an NP-hard optimization problem. To solve this problem, we
propose a low-complexity approach which is based on sequentially optimizing
each entry in $\mathcal{P}^{\left(i\right)}$, i.e. $p_{1}^{\left(i\right)}$,
$p_{2}^{\left(i\right)}$,..., $p_{Q}^{\left(i\right)}$, one-by-one.
Specifically, we define $\mathcal{P}_{q}^{\left(i\right)}=\left\{ p_{1}^{\left(i\right)},...,p_{q}^{\text{\ensuremath{\left(i\right)}}},p_{q+1}^{\left(i-1\right)},p_{q+2}^{\left(i-1\right)},...,p_{Q}^{\left(i-1\right)}\right\} \subset\mathcal{S}$
as the intermediate state of feeding port indices at the $i$th iteration
where the indices of the first $q$ feeds have been optimized and
the remaining $Q-q$ feeds not optimized. When optimizing index $p_{q}^{\left(i\right)}$
for the $q$th feeding port, we fix the other $Q-1$ indices and select
the optimal $p_{q}^{\left(i\right)}$ from the remaining $S-Q+1$
indices in $\mathcal{S}$ to minimize the correlation coefficient
\eqref{eq:CorrCoef} among all $Q$ ports in $\mathcal{P}_{q}^{\left(i\right)}$,
which can be formulated as
\begin{align}
\underset{p_{q}^{\left(i\right)}}{\mathrm{min}}\:\:\:\: & \stackrel[k\in\mathcal{P}_{q}^{\left(i\right)}]{j\neq k}{\sum}\stackrel[j\in\mathcal{P}_{q}^{\left(i\right)}]{}{\sum}\left|\rho_{jk}\left(\mathrm{\mathbf{X}}^{L\left(i\right)}\right)\right|,\label{eq:optimum ports update}\\
\mathrm{s.t.}\:\:\:\: & p_{q}^{\left(i\right)}\in\mathcal{S}\setminus\left(\mathcal{P}_{q-1}^{\left(i\right)}\setminus p_{q}^{\left(i-1\right)}\right),\label{eq:optimum ports update constraint}
\end{align}
with $q$ sequentially taken as $q=1,2,...,Q$. It should be noted
that $\mathcal{P}_{q-1}^{\left(i\right)}=\mathcal{P}^{\left(i-1\right)}$
when $q=1$.
\begin{algorithm}[t]
\caption{The alternating optimization method\label{alg:AO}}

\textbf{Input:} $\mathcal{P}^{\text{\ensuremath{\left(0\right)}}}$,
$\mathbf{X}^{L\mathrm{\left(0\right)}}$ (initial values described
in text);

$\quad${\scriptsize{}1:} \textbf{Initialization:} $i=0$;

$\quad${\scriptsize{}2:} \textbf{repeat}

$\quad$$\quad$$\:\:\:\:$$i=i+1$;

$\quad${\scriptsize{}3:} $\:\:\:\:$Find $\mathbf{X}^{L\left(i\right)}$
with $\mathcal{P}^{\text{\ensuremath{\left(i-1\right)}}}$ by \eqref{eq:optimum};

$\quad${\scriptsize{}4:} $\:\:\:\:$\textbf{for} $q=1\colon Q$

$\quad${\scriptsize{}5:} $\:\:\:\:$$\:\:\:\:$Find $p_{q}^{\left(i\right)}$
subject to \eqref{eq:optimum ports update constraint} with $\mathbf{X}^{L\left(i\right)}$
by \eqref{eq:optimum ports update};

$\quad${\scriptsize{}6:} $\:\:\:\:$$\:\:\:\:$Update $\mathcal{P}_{q}^{\left(i\right)}$;

$\quad${\scriptsize{}7:} $\:\:\:\:$\textbf{end}

$\quad${\scriptsize{}8:} $\:\:\:\:$\textbf{$\mathcal{P}^{\left(i\right)}=\mathcal{P}_{Q}^{\left(i\right)}$}

$\quad${\scriptsize{}9:} \textbf{until} $\mathcal{P}^{\left(i\right)}=\mathcal{P}^{\left(i-1\right)}$;

\textbf{Output:} $\mathbf{X}^{L\star}=\mathbf{X}^{L\left(i\right)}$,
$\mathcal{P}^{\star}=\mathcal{P}^{\left(i\right)}$;
\end{algorithm}

By iteratively optimizing the load reactances and selecting feeding
ports, the objective function \eqref{eq:optimum} and \eqref{eq:optimum ports update}
can converge to a local optimal solution \cite{bezdek2003convergence}.
Therefore, we obtain the optimal load reactances $\mathbf{X}^{L\star}$
of the loaded $N$-port structures with $Q$ feeds with indices of
$\mathcal{P}^{\star}$. Algorithm \ref{alg:AO} summarizes the overall
algorithm for optimizing the load reactances and feeding ports. The
performance of the algorithm in finding optimum MIMO antenna configuration
will be shown in the next section.

\section{Numerical Results\label{sec:Numerical-Simulation}}

In this section, we firstly compare our proposed method with previous
work \cite{jensen2008capacity} to verify its accuracy and correctness.
After that we propose a MIMO antenna that can approach the continuous
channel capacity bound.

For channel capacity simulations with the current constraint, we use
the same SNR definition as in \cite{jensen2008capacity} which is
the ratio of the current norm $I_{\mathrm{in}}^{2}$ to the noise
power $\sigma^{2}$. While for capacity simulation with the radiated
power constraint and also dual constraints, we use the conventional
SNR definition, i.e. the ratio of maximum radiated power $P_{\mathrm{rad}}$
to the noise power $\sigma^{2}$.

\subsection{Comparison of Channel Capacity}

\begin{figure}[t]
\begin{centering}
\textsf{\includegraphics[width=6cm]{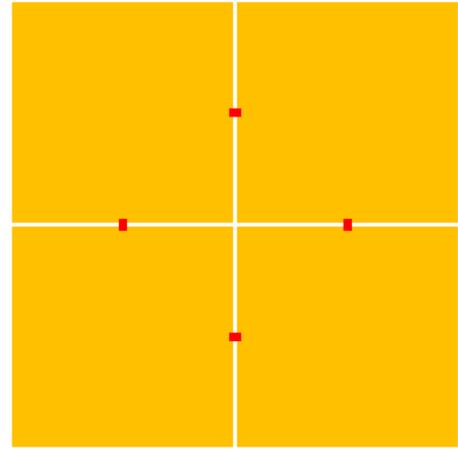}}
\par\end{centering}
\begin{centering}
\textsf{\includegraphics[width=6cm]{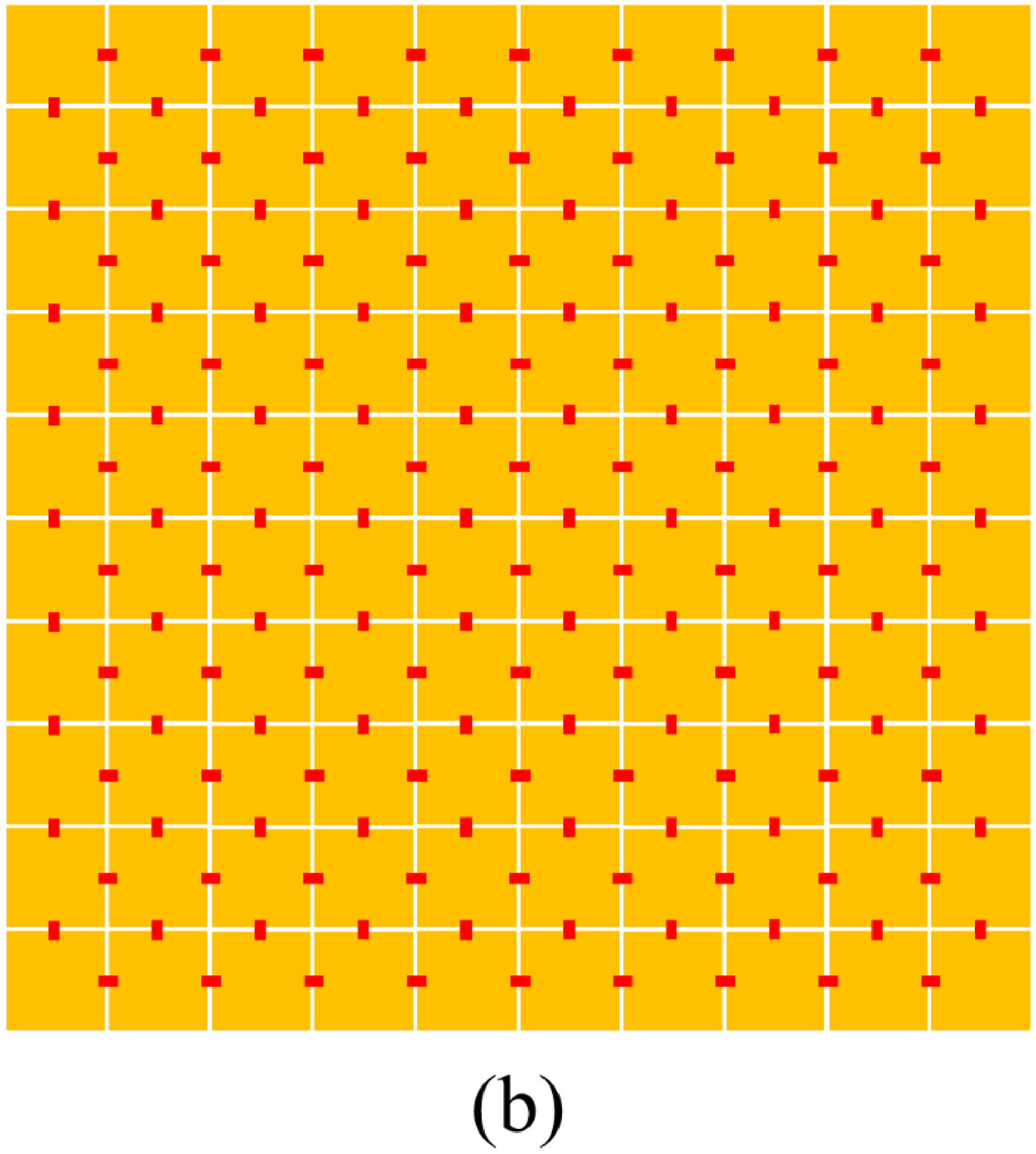}}
\par\end{centering}
\caption{Discretization of a one square wavelength 2D square planar surface
using an $N$-port loaded structure where the ports are indicated
by the red marks. In example a) there are $2\times2$ sub-elements
with 4 ports, and for example b) there are $10\times10$ sub-elements
with 180 ports. \label{fig:Jensen}}
\end{figure}
In the first set of simulation results we wish to verify our approach
by comparing with previous results \cite{jensen2008capacity}. The
previous results consider a continuous source current on a 2D square
planar surface with one square wavelength size at both the transmitter
and receiver. The 2D surface is discretized into an $N$-port structure
as shown in Fig \ref{fig:Jensen} where discretizations with 4 and
180 ports are shown. This corresponds to dividing the one wavelength
square surface into 4 and 100 sub-elements respectively similarly
to the discretizations of previous work \cite{jensen2008capacity}.

To compare our method to previous work \cite{jensen2008capacity},
we only need to find the required currents at all $N$ ports and do
not need to consider finding $N-Q$ loads or $Q$ feeding ports. We
can therefore use the results in \eqref{eq:optimized capacity-1-1}
and \eqref{eq:optimized capacity constraint-1-1} directly without
considering the loads and feeds to find the port currents.

The same channel setup utilized previously \cite{jensen2008capacity}
is also invoked and includes single polarization in a rich scattering
environment with Rayleigh fading and 2D uniform power angular spectrum
(PAS) on the azimuth plane \cite{PCDM}. In addition, we use the current
norm constraint so that the capacity optimization problem is formulated
as \eqref{eq:optimized capacity-1-1} and \eqref{eq:optimized capacity constraint-1-1},
with WF to find the currents. This is similar to equation (21) in
\cite{jensen2008capacity}.

\begin{figure}[t]
\begin{centering}
\textsf{\includegraphics[width=8.5cm]{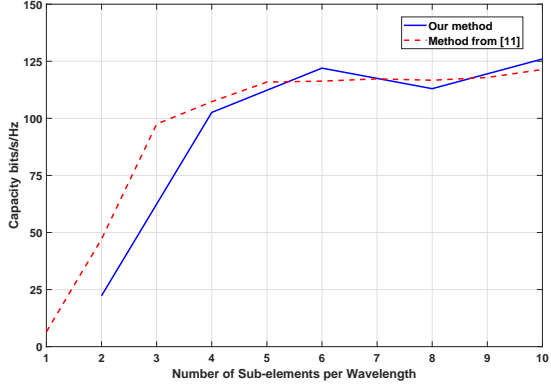}}
\par\end{centering}
\caption{Capacity comparison with method in \cite{jensen2008capacity} at $\mathrm{SNR}=20$
dB. Note that the discretizations on the horizontal axis are sub-elements
per wavelength. That is 10 sub-elements per wavelength corresponds
to 100 sub-elements in total.\label{fig:Cap compare}}
\end{figure}

Simulation results of the system capacity are shown in Fig. \ref{fig:Cap compare}
for various levels of discretizations. In the figure the number of
discretizations per dimension is used to align with previous work
\cite{jensen2008capacity}. That is 10 sub-elements per wavelength
correspond to 100 sub-elements in total. It can be observed that by
using our method, the system capacity approaches that using the method
in \cite{jensen2008capacity} when the number of sub-elements per
wavelength is more than four. The capacity gap (for 4 or less sub-elements
per square wavelength) arises due to the current on the discrete ports
being an approximation to the continuous current on the surface, which
is related to the position of the ports in the $N$-port structure.
Therefore, we can conclude that our method can obtain the continuous-space
electromagnetic channel capacity bound if the dimension of the sub-element
and the separation distance between ports is less than a wavelength
(from Fig. \ref{fig:Jensen} greater than 5 sub-elements per wavelength).

\subsection{MIMO Antenna Capacity Bound}

Next, we use the technique in Section IV to design a practical MIMO
antenna that approaches the capacity bounds. The frequency considered
in the following is 2.4 GHz so that the wavelength is 125 mm.

\begin{figure}[t]
\begin{centering}
\textsf{\includegraphics[width=8.5cm]{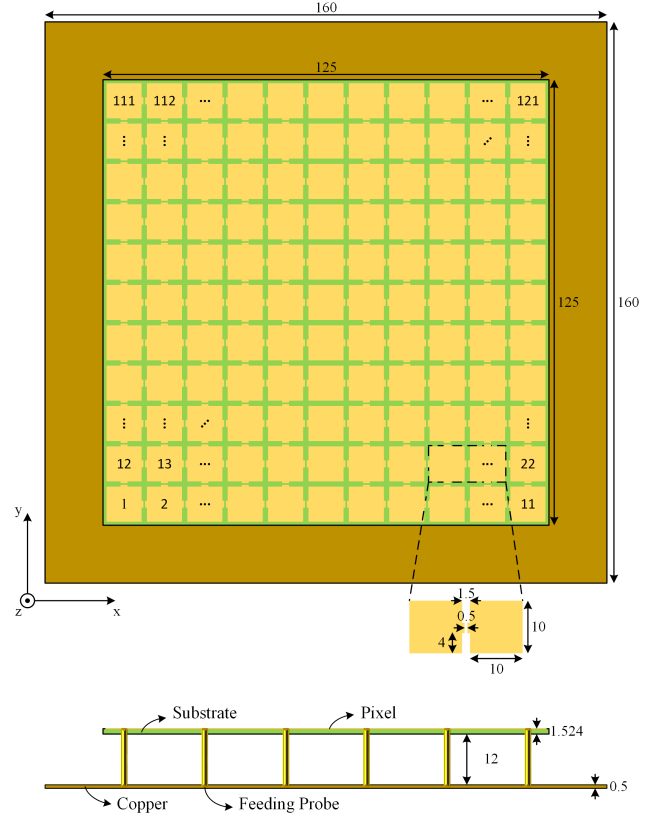}}
\par\end{centering}
\caption{Plan and elevation views of the proposed $N$-port structure. A 2D
square copper surface with one square wavelength size on top of a
ground plane is utilized.\label{fig:Plane}}
\end{figure}

The first step is to propose a structure that is implementable with
$Q$ feeds but general enough to allow the formation of arbitrary
current distributions for capacity maximization. As such our proposed
$N$-port structure is shown in Fig. \ref{fig:Plane} where a 2D square
copper planar surface with one square wavelength size is again utilized.
The copper surface is mounted on a substrate where the copper has
electric conductivity of $5.8\times10^{7}$ S/m while the substrate
is made with Rogers 5880C which has permittivity of 2.2 and loss tangent
of 0.0009. We discretize the 2D copper surface into a $11\times11$
sub-element array where the size of the sub-element is $10\times10\:\mathrm{mm}^{2}$.
To implement the feeds, an additional copper plane is placed underneath
as a ground plane so that a conventional feed can be connected between
the ground and the center of a sub-element as shown in the elevation
view in Fig. \ref{fig:Plane}. The $N$ ports of the structure are
defined as the ports across each pair of adjacent sub-elements on
the surface and also between the ground and the center of each sub-element.
That is there are 220 ports on the surface and 121 ports between the
ground and sub-elements making up a total of $N=341$ ports. The $Q$
feeding ports need to be selected from the 121 ports between the ground
and sub-elements.

While the ultimate aim is to select $Q$ feeding ports from the ground
to sub-element ports, we firstly need to find the required currents
on the $N$ ports to optimize capacity and approach the capacity bounds
similarly as in Section \ref{sec:Numerical-Simulation}.A. This allows
us to first determine the capacity performance of the structure with
ideal port currents. The final load reactances and feeds arrangements
of the antenna that can approximate these currents are found in the
next subsection.

In the capacity simulations, we obtain them as a function of $P$
($P=1,2,3$, etc.) basis with the highest eigenvalues (out of $N=341)$
for transmission. This allows us to determine the number of EADoF
provided by the structure. In the simulations, we use $P$ ideally
isolated antennas at the receiver side. We also assume a rich scattering
environment so that entries in $\mathbf{H}_{v}$ satisfy $\left[\mathbf{H}_{v}\right]_{ij}\sim\mathcal{CN}\left(0,1\right)$
for $i,j=1,2,\ldots,K$. In addition, dual polarization in a rich
scattering environment with Rayleigh fading and 3D uniform PAS over
full sphere is utilized. For these reasons the capacity results obtained
for this channel will be much greater than that obtained in Section
V.A which used only one polarization, a 2D PAS and a receiver antenna
that was the same as the transmitter.

\subsubsection{Capacity Bound with Individual Constraints}

\begin{figure}[t]
\begin{centering}
\textsf{\includegraphics[width=8.5cm]{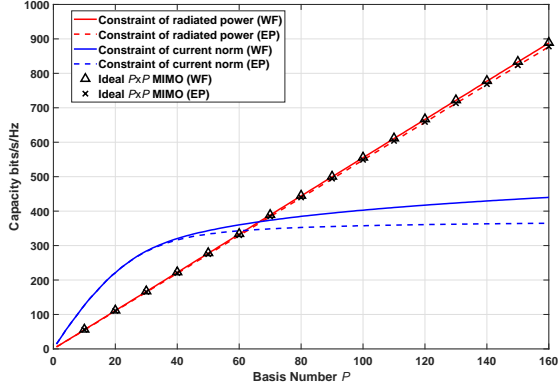}}
\par\end{centering}
\caption{Capacity of $P\times P$ MIMO system using the $N$-port structure
in Fig. \ref{fig:Plane} with the constraints of radiated power and
current norm at $\mathrm{SNR}=20$ dB. \label{fig:Cap Cons 1 2}}
\end{figure}
We simulate the capacity of the proposed structure when we use the
$P$ basis functions with the highest eigenvalues to create a $P\times P$
system with the two individual constraints of current norm and radiated
power. The simulated capacity with the two constraints using EP allocation
and WF method are shown in Fig. \ref{fig:Cap Cons 1 2}. These are
benchmarked with the results of an ideal $P\times P$ MIMO system
where the transmitter is equipped with $P$ spatially isolated antennas.
It can be observed that if we only consider the constraint of radiated
power, the simulated capacity of the structure is the same as that
of ideal MIMO. However, the current norm in this case can be extremely
large and not implementable.

On the other hand, if we only consider the constraint of current norm,
the capacity of the loaded structure is higher than the ideal MIMO
systems for the first few basis. This is because those basis with
large eigenvalues have larger radiated power than the ideal MIMO system.
However, when we consider more basis, e.g. more than 50, the increase
in capacity is negligible since those basis have small eigenvalues
and only radiate limited power. Therefore, in this case the WF method
can allocate more power to these 50 basis with large eigenvalues to
improve the capacity.

\subsubsection{Capacity Bound with Dual Constraint}

\begin{figure}[t]
\begin{centering}
\textsf{\includegraphics[width=8.5cm]{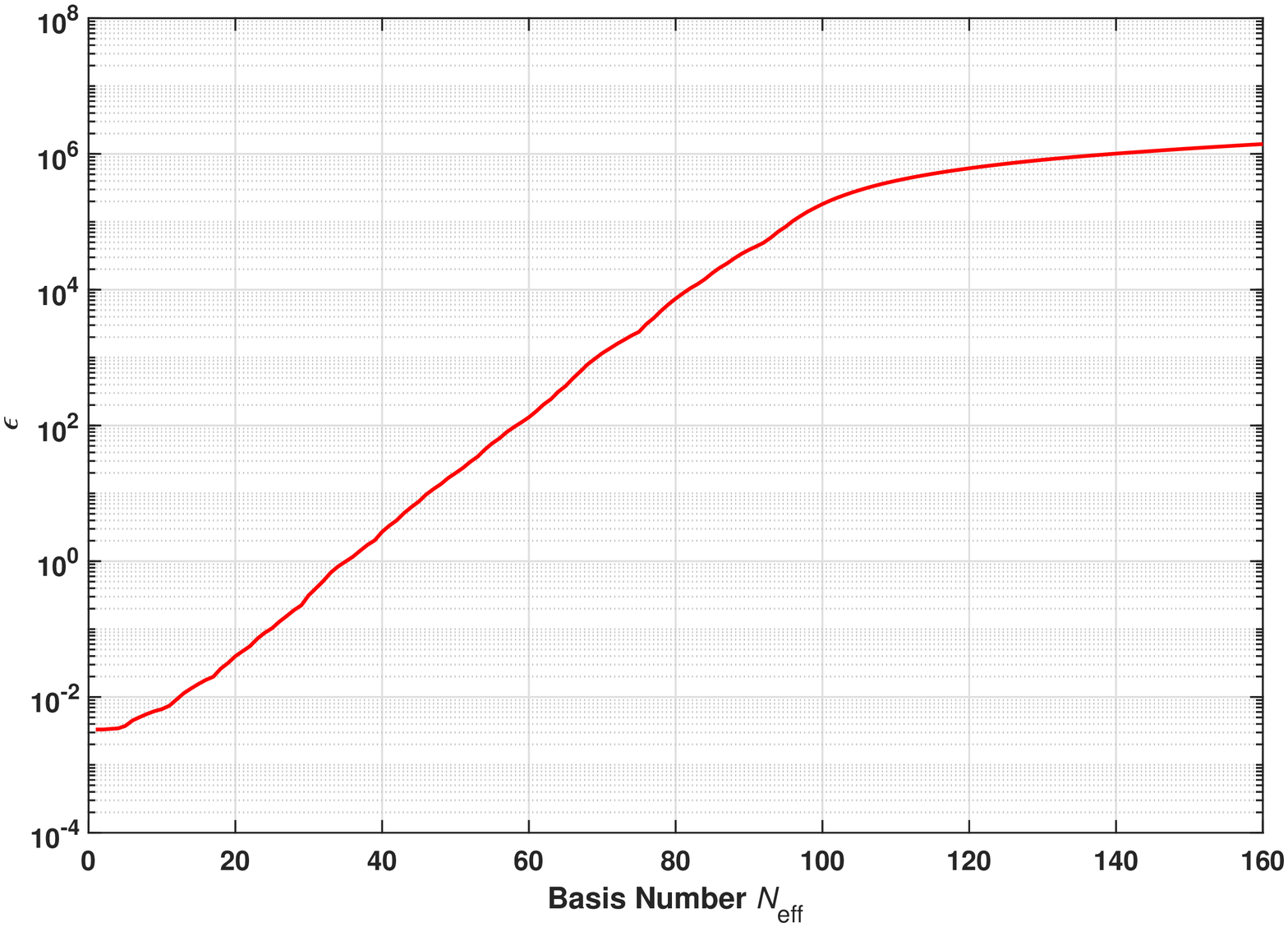}}
\par\end{centering}
\caption{Results relating the dual constraints ratio, $\epsilon$ to the EADoF
$N_{\mathrm{eff}}$. For $\epsilon=0.02$ it can be observed $N_{\mathrm{eff}}\approx18.$
\label{fig:epsilon}}
\end{figure}

Next we consider imposing dual constraints for the simulation of channel
capacity and this leads us to consider $\epsilon$ and $N_{\mathrm{eff}}$.
In Fig. \ref{fig:epsilon}, we plot the minimum $\epsilon$ for the
first $N_{\mathrm{eff}}$ basis as calculated in \eqref{eq:EP for Dual Cons}.
It can be noticed that $\epsilon$ is an increasing function of $N_{\mathrm{eff}}$
as expected.

For an ideal lossless antenna, the input impedance is $Z_{\mathrm{in}}=50\ \Omega$,
i.e. $P_{\mathrm{rad}}=Z_{\mathrm{in}}I_{\mathrm{in}}^{2}$. Therefore
a region of interest on Fig. \ref{fig:epsilon}, is when $\epsilon=I_{\mathrm{in}}^{2}/P_{\mathrm{rad}}=1/Z_{\mathrm{in}}=0.02$.
In this region $N_{\mathrm{eff}}$ is approximately 18 and therefore
provides an estimate of the minimum EADoF to expect. In essence we
can expect to achieve at least 18 ports in the one square wavelength
area. We consider it as a lower bound in practice, depending on the
exact relation between $I_{\mathrm{in}}^{2}$ and $P_{\mathrm{rad}}$.
We may be able to achieve more ports because the internal currents
in the antenna can be higher than those at the feeds.

\begin{figure}[t]
\begin{centering}
\textsf{\includegraphics[width=8.5cm]{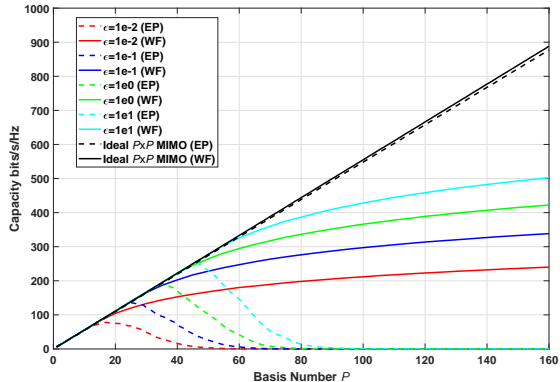}}
\par\end{centering}
\caption{Capacity of $P\times P$ MIMO using using the $N$-port structure
in Fig. \ref{fig:Plane} with dual constraints at $\mathrm{SNR}=20$
dB.\label{fig:Capacity Dual}}
\end{figure}

In Fig. \ref{fig:Capacity Dual} we plot the resulting capacity when
we use the first $P$ basis functions to create a $P\times P$ system
for various $\epsilon$ using EP and WF. It can be observed again
that the capacity of the MIMO system using the $N$-port structure
increases as $\epsilon$ becomes larger. When considering basis $P$
smaller than $N_{\mathrm{eff}}$, the system capacity of MIMO using
the loaded structure is the same as that of ideal $P\times P$ MIMO.
However when $P$ approaches $N_{\mathrm{eff}}$, the capacity using
EP allocation starts to decrease since equality in dual constraints
cannot be achieved. The WF method allocates most power to the first
$N_{\mathrm{eff}}$ basis so that the capacity slightly increases
when adding more basis to the $P\times P$ MIMO system. In addition,
the $\epsilon$ value points where the WF and EP lines separate indicate
the number of available basis $N_{\mathrm{eff}}$ (i.e. EADoF) provided
by the $\epsilon$ value structure as calculated in \eqref{eq:EP for Dual Cons}
and approximately match with the results in Fig. \ref{fig:epsilon}.

The results in Fig. \ref{fig:epsilon} and and Fig. \ref{fig:Capacity Dual}
are useful references in designing MIMO antenna for approaching the
capacity bound. The use of $\epsilon$ provides us with an estimate
of the number of feeding ports or $N_{\mathrm{eff}}$. In particular,
we can expect to be able to provide at least 18 ports in the one square
wavelength size antenna structure shown in Fig. \ref{fig:Plane}.

\subsection{MIMO Antenna Design}

To provide a useful antenna, we need to select feeding ports and the
loads for all the ports in the $N$-port loaded structure. The feeding
ports and loads must be selected to achieve a close match to the optimum
currents found in the previous section for a feasible $P$. This then
would provide the optimum antenna configuration that can achieve the
channel capacity bound with dual constraints.

As discussed previously for an ideal lossless antenna, the input impedance
is $Z_{\mathrm{in}}=50\ \Omega$ and therefore $\epsilon\geqslant I_{\mathrm{in}}^{2}/P_{\mathrm{rad}}=1/Z_{\mathrm{in}}=0.02$
so that $N_{\mathrm{eff}}$ is at least 18. Expecting to have higher
currents internally on the antenna, we aim for 20 ports in our design.
That is we will need to re-create the currents obtained when $P=20$
from the previous section.

To reduce the computational effort in finding the loads, we preset
the sub-elements in the center row and center column of the 2D copper
surface as being connected together (shorted) so there is a single
large cross element centered on the surface. That is ports \{56, 57,
58, 59, 60, 61, 62, 63, 64, 65, 66, 6, 17, 28, 39, 50, 72, 83, 94,
105, 116\} in Fig. \ref{fig:Plane} are shorted together. The cross
element separates the other sub-elements into four sub-square arrays.
The cross is also isolated from the other sub-elements so that the
ports between the cross and the other adjacent sub-elements are open
without loads. In effect, we are presetting some of the ports to be
shorted or open on the surface. This reduces the computational load
of the optimization by reducing the number of load reactances to be
found from 220 to 160. In addition, to reduce the search space for
the $Q=20$ feeds we also restrict it to the $S=64$ ports between
the ground plane and the sub-elements on the edge of each sub-square
array (instead of all 121 possible locations). For those ports underneath
the surface that were not selected as feeds, we set them to be open.

\begin{figure}[t]
\begin{centering}
\textsf{\includegraphics[width=8.5cm]{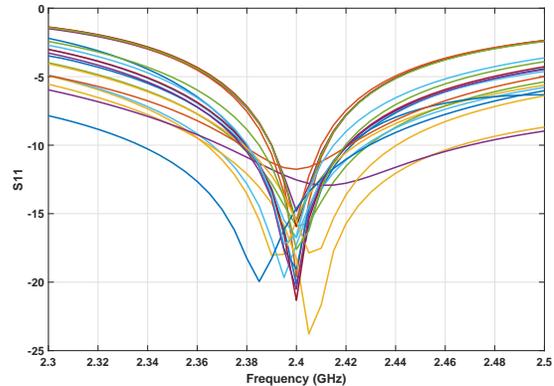}}
\par\end{centering}
\caption{Return loss, S11, of the 20 feeding ports in the loaded structure
in Fig. \ref{fig:Plane}.\label{fig:20 ports S11}}
\end{figure}

To find the optimum reactances across the 160 ports on the surface
and the indices of the 20 feeding ports underneath, we use the method
as described in Section \ref{sec:Antenna Design} so that the excited
radiation patterns of the 20 feeding ports are closest to being orthogonal
\eqref{eq:optimum}. The final indices of the 20 feeding ports after
the alternating optimization are given by

{\footnotesize{}
\[
\left\{ 1,3,5,7,9,11,29,33,49,55,67,69,75,89,95,111,113,117,119,121\right\} 
\]
}and their physical location can be found from Fig. \ref{fig:Plane}.
The 20 feeding ports are then each fed by a feeding probe across the
common ground and the center of the selected sub-elements as shown
in the elevation view in Fig. \ref{fig:Plane}. The impedance matching
for the 20 feeding ports is performed separately with a conventional
T-type matching network to achieve matching to 50 $\Omega$. The return
loss (S11) results of the 20 ports are provided in Fig. \ref{fig:20 ports S11}
where the reflected power at 2.4 GHz is around -15 dB and the bandwidth
is around 40 MHz.

\begin{figure}[t]
\begin{centering}
\textsf{\includegraphics[width=8.5cm]{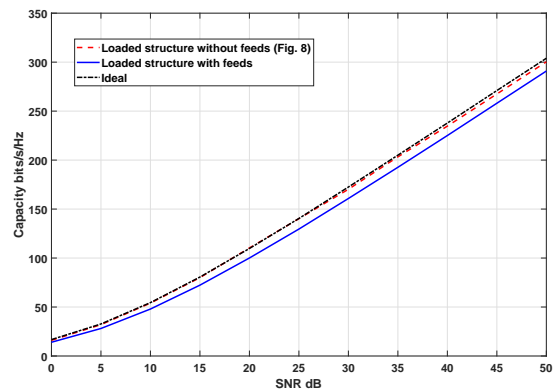}}
\par\end{centering}
\caption{Capacity of $20\times20$ MIMO system for the loaded structure in
Fig. \ref{fig:Plane} with and without feeds.\label{fig:20 ports MIMO Capacity}}
\end{figure}

By using the 20 excited radiation patterns from the resultant loaded
structure with 20 feeds, and a beamspace channel model \eqref{eq:angular MIMO system},
the capacity of the $20\times20$ MIMO system is found and the simulated
results are shown in Fig. \ref{fig:20 ports MIMO Capacity}. These
are also benchmarked with the capacity bound of the loaded structure
without feeds obtained by WF method in Section III.C and the capacity
of ideal MIMO. It can be observed that the MIMO antenna using the
loaded structure with 20 feeds achieves performance close to ideal
MIMO and the capacity bound predicted by Fig. \ref{fig:Capacity Dual}
for loaded structure without feeds. The capacity gap is due to the
small correlations among the radiation patterns of the feeding ports
and power loss from mutual coupling. However, the key advantage of
the proposed method is that the optimum antenna structure for the
MIMO transmitter with size constraint can be found with only minor
compromise on system capacity.

\section{Conclusion}

In this paper, we have proposed a novel method for designing antennas
that approach the continuous-space electromagnetic channel capacity
bounds. The method can link continuous-space electromagnetic channel
capacity bounds to MIMO antenna design. The method is not restricted
to a specific antenna configuration and uses loaded $N$-port structures
to discretize the continuous space and represent arbitrary antenna
geometries. It is useful for designing compact MIMO antennas that
can approach channel capacity bounds when constrained by size.

In the proposed method, we derive the closed-form expressions for
the channel capacity limits using a beamspace channel model with a
current constraint, radiated power constraint as well as with dual
constraints. We also introduce a method for antenna design using the
loaded ports structure and provide an efficient alternating optimization
approach for finding the optimum MIMO antenna configuration to generate
orthogonal radiation patterns. This can be used to construct a beamspace
MIMO system that approaches the channel capacity bounds.

Simulation results of the channel capacity using our proposed method
matches well with previous work. Furthermore, by optimizing the load
reactance and feeding port positions in our proposed antenna design
with one square wavelength size, the achieved capacity performance
is close to the continuous-space channel capacity bounds, demonstrating
the effectiveness of the proposed method. In particular we show that
at least 18 ports can be supported in a one wavelength square structure
to achieve the continuous-space electromagnetic channel capacity bound.
It is also shown that the limit on the number of ports is constrained
by the maximum current that the antenna can handle. An example design
for a 20-port antenna in a one square wavelength area that achieves
the capacity bounds is also provided. One challenge of the final antenna
is the limited bandwidth and this can be addressed by increasing the
height in the antenna design.

\end{document}